\documentstyle[11pt]{article}
\input pictex.sty
\begin{document}
\font\thinlinefont=cmr5
\vspace*{-2cm}
\noindent
\hspace*{11cm}
UG--FT--62/96 \\
\hspace*{11cm}
hep-ph/9605418 \\
\hspace*{11cm}
May 1996 \\
\begin{center}
\begin{large}
{\bf Invariant formulation of CP violation \\
for four quark families \\}
\end{large}
~\\
F. del Aguila, J. A. Aguilar--Saavedra \\
{\it Departamento de F\'{\i}sica Te\'{o}rica y del Cosmos \\
Universidad de Granada \\
18071 Granada, Spain}
\end{center}
\begin{abstract}
We find a minimal set of constraints which are independent of the
choice of weak quark basis and necessary and sufficient for CP
conservation for four quark families, including also the case
of degenerate quark masses. These invariant conditions are written
in the mass eigenstate basis as a function of the fermion masses and
charged current mixings. CP violation is then related to the areas of
three unitarity quadrangles and the CP violating effects of the fourth
family are discussed in the case of small mixings. 
\end{abstract}
\hspace{0.8cm}
Pacs: 11.30.Er, 12.15.Ff, 14.65.-q, 14.80.-j
\section{Introduction.}
The observed CP violation in the $\mathrm K^0$-$\mathrm \bar K^0$
system is accounted for in the standard model by the 
Cabibbo-Kobayashi-Maskawa (CKM) matrix \cite{papiro1}. There is only
one CP violating phase for three families of quarks. If a fourth family
exists, there are three CP violating phases \cite{nose1}.
Although the $Z^0$ invisible width excludes a fourth generation with
a light neutrino \cite{papiro1a}, an extra heavy family is not ruled out
\cite{papiro1b}. In fact, the deviation from the pattern of CP
violation expected for three families, for instance, in b physics may
signal to a fourth quark generation \cite{papiro1c}.

\begin{figure}[htb]
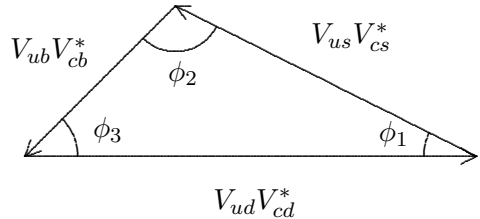

\begin{center}
\mbox{\beginpicture
\setcoordinatesystem units <1cm,1cm>
\unitlength=1cm
\linethickness=1pt
\setplotsymbol ({\thinlinefont .})
\plot 0 0 6 0 /
\plot 6 0 2 2 /
\plot 2 2 0 0 /
\plot 5.8 0.1 6 0 5.8 -0.1 /
\plot 2.2 2 2 2 2.134 1.822 /
\plot 0.0707 0.212 0 0 0.212 0.0707 /
\circulararc 45 degrees from 0.7 0 center at 0 0
\circulararc -26.56 degrees from 5.3 0 center at 6 0
\circulararc 108.43 degrees from 1.57 1.57 center at 2 2
\put{$V_{ud} V_{cd}^*$} [lB] at 2.5 -0.7
\put{$V_{us} V_{cs}^*$} [lB] at 3.8 1.5
\put{$V_{ub} V_{cb}^*$} [lB] at -0.2 1.3
\put{$\phi_3$} [lB] at 0.9 0.25
\put{$\phi_2$} [lB] at 1.9 1.0
\put{$\phi_1$} [lB] at 4.7 0.2
\linethickness=0pt
\putrectangle corners at -1 -1 and 7 3
\endpicture}
\end{center}
\caption{Unitarity triangle \label{figura1} }
\end{figure}

The CKM matrix is defined in the quark mass eigenstate basis and
therefore, it is
not invariant under arbitrary unitary transformations of the quark
basis. For three fermion families it has been identified a quantity
invariant under quark basis transformations whose vanishing
characterizes CP conservation \cite{papiro2}:
\begin{equation}
I \equiv \det\; [M_u M_u^\dagger,M_d M_d^\dagger]=0,
\label{eq:1:1}
\end{equation}
where $M_{u,d}$ are the up and down mass matrices. This invariant
formulation makes apparent the necessary and sufficient conditions
for CP conservation ({\em i\/}). Thus
\begin{eqnarray}
I & = & -2 i (m_t^2-m_c^2)(m_t^2-m_u^2) (m_c^2-m_u^2)  
 (m_b^2-m_s^2)(m_b^2-m_d^2)(m_s^2-m_d^2) \nonumber \\
& \times & {\mathrm Im\;}
 (V_{ud} V_{cd}^* V_{cs} V_{us}^{*}), \label{eq:1:2}
\end{eqnarray}
where $m_i$ is the mass of the quark $i$ and $V_{ij}$ is the $ij$
entry of the CKM matrix, implies that CP is conserved if some up or
down quark masses are degenerate or a product
$V_{ij} V_{kj}^* V_{kl} V_{il}^*$, $i\neq j$, $k\neq l$ is real. 
On the other hand, any CP
violating observable is proportional to the factors in Eq.
(\ref{eq:1:2}) which then give the size of CP violation ({\em ii\/}).
 This
invariant formulation also allows to decide in any weak basis if CP
is conserved ({\em iii\/}). It also motivates model building ({\em
iv\/}) and it can eventually
help to understand the origin of CP violation if for some reason a
definite model (weak basis) is physically distinguished ({\em v\/}).
 In three dimensions
\cite{papiro3}
\begin{equation}
{\mathrm tr\;} [M_u M_u^\dagger,M_d M_d^\dagger]^3=3I.
\label{eq:1:3}
\end{equation}
Hence this trace is also imaginary (see Eq. (\ref{eq:1:2}) ) and
proportional to the area $A$ of a triangle with sides $V_{ud}
V_{cd}^*$, $V_{us} V_{cs}^*$, $V_{ub} V_{cb}^*$ and angles
$\phi_{1-3}$
with 
\begin{eqnarray*}
\sin \phi_1 & = &|\sin \arg\,(V_{ud} V_{cd}^* V_{cs} V_{us}^*)|, \\
\sin \phi_2 & = &|\sin \arg\,(V_{us} V_{cs}^* V_{cb} V_{ub}^*)|, \\
\sin \phi_3 & = &|\sin \arg\,(V_{ub} V_{cb}^* V_{cd} V_{ud}^*)|
\end{eqnarray*}
in the complex plane (see Fig. \ref{figura1}) \cite{papiro4,papiro8},
\begin{equation}
{\mathrm tr\;} [M_u M_u^\dagger,M_d M_d^\dagger]^3 \propto {\mathrm
Im\;} (V_{ud} V_{cd}^{*} V_{cs} V_{us}^{*})=2 A.
\label{eq:1:4}
\end{equation}
Hence the vanishing of any angle (side) of this triangle stands for
CP conservation. A b factory 
will help to determine the triangle better and to verify if the 
pattern of CP
violation corresponds to the existence of three families only
\cite{papiro9}.  (In
this case other equivalent triangles resulting from the unitarity of the
CKM matrix and involving the $b$ quark may be more convenient
\cite{papiro1b,papiro1c,papiro9}.)

In this paper we revise the analogous formulation for four quark
families. There is a large literature on the subject and many of our
results have been first obtained by other authors
\cite{papiro8,papiro6}. We
have taken advantage of the extensive use of symbolic programs as 
{\em Mathematica\/} \cite{papiro7} for the detailed proofs and for obtaining
explicit results \cite{papiro7b}. Very often it was necessary educated guessing to
avoid otherwise unmanageable intermediate expressions. When this was
possible it could be traced back to a deeper, sometimes already known
reason. A set of conditions independent of the choice of weak
quark basis and necessary and sufficient for CP conservation for four
generations was found for nondegenerate quark masses in Ref.
\cite{papiro6}. First we extend this set to also include the case of
degenerate quark masses. Afterwards we write these invariant quantities
as a function of physical parameters, {\em i.e.\/} quark masses and
products of CKM matrix elements independent of the choice of the
phases of quark mass eigenstates. Then the requirement that the area of
the triangle in Eq. (\ref{eq:1:4}) vanishes if CP is conserved for
three families is generalized to the vanishing of the areas of
three quadrangles for four generations \cite{papiro8}. 
Finally we comment on the realistic case of small mixings.

\section{Complete set of invariant constraints for CP conservation.}
In the standard model with $N$ families the quark mass term can be
written, after spontaneous symmetry breaking,
\begin{equation}
-{\mathcal L}_{mass}=\bar u_L M_u u_R + \bar d_L M_d d_R + 
{\mathrm h.c.},
\label{eq:2:1}
\end{equation}
where $u_{L,R}$, $d_{L,R}$ are weak eigenstates. This Lagrangian is
invariant under a CP transformation leaving the $\mathrm SU(2)_L \times U(1)_Y$
gauge interactions unchanged \cite{papiro3} 
\begin{eqnarray}
u_L\rightarrow U_L C u_L^* &~,~~& u_R\rightarrow U_R^u C u_R^*~, \nonumber \\
d_L\rightarrow U_L C d_L^* &~,~~& d_R\rightarrow U_R^d C d_R^* ~,
\end{eqnarray}
where $C$ is the Dirac charge-conjugation matrix and 
$U_L$, $U_R^{u,d}$ are $4 \times 4$ unitary matrices, if
\begin{equation}
U_L^\dagger M_u U_R^u=M_u^*~,~~~
U_L^\dagger M_d U_R^d=M_d^*~.
\label{eq:2:3}
\end{equation}
Thus CP is conserved if $U_L$, $U_R^{u,d}$ exist fulfilling Eq.
(\ref{eq:2:3}). In the standard model with only one Higgs doublet
$U_R^{u,d}$ are unobservable. Hence, as an
arbitrary complex matrix can be written as the product of a hermitian 
matrix with
non-negative eigenvalues times a unitary matrix, $M_{u,d}$ can be
assumed to be hermitian with non-negative eigenvalues. Then,
\begin{equation}
U_L^\dagger H_u U_L=H_u^*~,~~~
U_L^\dagger H_d U_L=H_d^*,
\label{eq:2:4}
\end{equation}
where $H_u=M_u M_u^\dagger$, $H_d=M_d M_d^\dagger$, are also necessary and 
sufficient conditions for CP conservation.
Under unitary transformations of the left-handed fields the trace of
any product of matrices $H_{u,d}$ is invariant. Therefore, Eq.
(\ref{eq:2:4}) implies
\begin{equation}
{\mathrm Im~tr\;}(H_u^{p_1} H_d^{p_2} H_u^{p_3} \cdots H_d^{p_r})=0.
\label{eq:2:5}
\end{equation}
with $(p_1, \dots, p_r)$ an arbitrary sequence of positive integers.
Eqs. (\ref{eq:2:5}), which are independent of the choice of quark 
basis, are also necessary and sufficient conditions for CP invariance 
\cite{papiro6}. However, in practice one
is interested in finding a minimal subset of these necessary and
sufficient conditions. For $N=3$ there is one such condition necessary
and sufficient for CP conservation,
\begin{equation}
{\mathrm Im~tr\;}(H_u^2 H_d H_u H_d^2)=0,
\end{equation}
where ${\mathrm Im~tr\;}(H_u^2 H_d H_u H_d^2)=-\frac{1}{6}
{\mathrm Im~tr\;}[H_u,H_d]^3$ in Eq. (\ref{eq:1:3}). 
For $N=4$ there are six conditions for CP invariance
for nondegenerate quark masses, defined by the sequences (2,1,1,2),
(2,1,1,3), (2,2,1,3), (1,1,1,2,1,3), (3,1,1,2), (3,1,1,3) in
Eq. (\ref{eq:2:5}) \cite{papiro6}. For the degenerate case two more
constraints must be added. In the {\em Appendix\/} we prove that a 
minimal set of these constraints consists of
\begin{eqnarray}
I_1 & = & {\mathrm Im~tr\;} (H_u^2 H_d H_u H_d^2)=0, \nonumber  \\
I_2 & = & {\mathrm Im~tr\;} (H_u^3 H_d H_u H_d^2)=0,  \nonumber \\
I_3 & = & {\mathrm Im~tr\;} (H_u^4 H_d H_u H_d^2-H_u^3 H_d H_u^2
H_d^2)=0, \nonumber \\
I_4 & = & {\mathrm Im~tr\;} (H_u^5 H_d H_u H_d^2-H_u^4 H_d H_u^2 H_d^2
  +  H_u^3 H_d H_u^2 H_d H_u H_d)=0,  \nonumber \\
I_5 & = & {\mathrm Im~tr\;} (H_u^2 H_d H_u H_d^3)=0,  \nonumber \\
I_6 & = & {\mathrm Im~tr\;} (H_u^3 H_d H_u H_d^3)=0,  \nonumber \\
I_7 & = & {\mathrm Im~tr\;} (H_u^2 H_d H_u H_d^4-H_u^2 H_d^2 H_u
H_d^3)=0, \nonumber \\
I_8 & = & {\mathrm Im~tr\;} (H_u^2 H_d H_u H_d^5-H_u^2 H_d^2 H_u H_d^4
  +  H_u H_d H_u H_d^2 H_u H_d^3)=0. \label{eq:inv1}
\end{eqnarray}
The important terms in $I_{3,4,7,8}$ are the sequences (3,\- 1,\- 2,\- 2),
(3,1,2,1,1,1), (2,2,1,3), (1,1,1,2,1,3), respectively. The other
terms are added to obtain more compact expressions later. We order
$I_{1-8}$ conventionally by increasing number of $H_d$ factors
because we assume $H_u$ diagonal and $H_d$ hermitian (as it can be
done without loss of generality) and in our proof the expressions for
the first conditions are simpler with this choice. At any rate the full
set is symmetric under the interchange of $H_u$ and $H_d$. For the 
nondegenerate case the sets $I_{1-6}=0$
and $I_{1,2,5-8}=0$ are both enough to guarantee CP
conservation. The second set is essentially the same as in Ref.
\cite{papiro6}, whereas the first one results from the interchange of 
$H_u$ and $H_d$. (In Ref. \cite{papiro6} $H_u$ was assumed to be
hermitian and $H_d$ diagonal. Then it was natural to look at
the conditions with smaller number of $H_u$ factors.) Obviously, the
symmetry of both sets is due to the impossibility of distinguishing up
and down quarks in Eqs. (\ref{eq:2:1},\ref{eq:2:4}). That the
complete set is necessary and sufficient for CP conservation even for
degenerate quark masses must be proven. We obtain this set of
constraints looking for all the invariant
expressions with $p=p_1+p_2+\cdots+p_r \leq 9$ in Eq. (\ref{eq:2:5}) 
and not identically zero. Invariants related by the cyclic
property of the trace and/or the hermiticity of $H_{u,d}$ are
counted once. We find
one ($I_1$) with $p=6$; two ($I_{2,5}$) with $p=7$; six with $p=8$
including the two terms of $I_{3,7}$ and $I_6$; and fourteen with 
$p=9$,
including the three terms of $I_{4,8}$. Then we solve the vanishing
conditions with increasing $p$ till we have no CP violating solution
left. The set in Eq. (\ref{eq:inv1}) is minimal in the sense that any
other subset of constraints with lower $p$ is not sufficient to
guarantee CP conservation.
There are many other choices of complete sets. For
instance, for nondegenerate quark masses $I_4$ in the complete
set $I_{1-6}$ could be replaced by $I_7$ and
\begin{eqnarray}
I_9 & = & {\mathrm Im~tr\;} (H_u^4 H_d H_u H_d^3 - H_u^3 H_d H_u^2
H_d^3), \nonumber \\
I_{10} & = & {\mathrm Im~tr\;} (H_u^3 H_d H_u H_d^4 - H_u^3 H_d^2 H_u 
H_d^3), \nonumber \\
I_{11} & = & {\mathrm Im~tr\;} (H_u^4 H_d H_u H_d^4 - H_u^4 H_d^2 H_u
H_d^3 
 -  H_u^3 H_d H_u^2 H_d^4 + H_u^3 H_d^2 H_u^2 H_d^3) \label{eq:inv2}
\end{eqnarray}
(see below). The proofs of the completeness of these sets were
based in the properties of hermitian matrices in Ref. \cite{papiro6}
and in the power of symbolic programs to solve
explicitly the constraints here. These conditions satisfy points
({\em iii\/}), ({\em iv\/}), ({\em v\/}) for four families. In order
to discuss points ({\em i\/}), ({\em ii\/}) we have to write these
constraints in the mass eigenstate basis, {\em i.e.\/} as a
function of quark masses and charged current mixings.

\section{Invariant formulation of CP violation in the mass eigenstate
basis.}
If we write $H_u=diag\;(m_1^2,m_2^2,m_3^2,m_4^2)$, 
$H_d=V_{CKM}\; diag\; (n_1^2,n_2^2,n_3^2,n_4^2)\; V_{CKM}^\dagger$
where 1, 2, 3, 4 stand for $u$, $c$, $t$, $t'$ in $H_u$ and
$d$, $s$, $b$, $b'$ in $H_d$, and $V_{CKM}$ is the unitary matrix diagonalizing
$H_d$, Eqs. (\ref{eq:inv1}, \ref{eq:inv2}) read
\begin{eqnarray}
I_1 & = & \sum_{i<j,k<l} G(i,j;k,l),  \nonumber \\
I_2 & = & \sum_{i<j,k<l} (m_i^2+m_j^2+m_4^2)\;G(i,j;k,l),\nonumber \\
I_3 & = & \sum_{i<j,k<l} (m_i^4+m_j^4+m_4^4)\;G(i,j;k,l),\nonumber \\
I_5 & = & \sum_{i<j,k<l} (n_k^2+n_l^2+n_4^2)\;G(i,j;k,l),\nonumber \\
I_6 & = & \sum_{i<j,k<l} (m_i^2+m_j^2+m_4^2)\;(n_k^2+n_l^2+n_4^2)\;
G(i,j;k,l), \nonumber \\
I_9 & = & \sum_{i<j,k<l} (m_i^4+m_j^4+m_4^4)\;(n_k^2+n_l^2+n_4^2)\;
G(i,j;k,l), \nonumber \\
I_7 & = & \sum_{i<j,k<l} (n_k^4+n_l^4+n_4^4)\;G(i,j;k,l),\nonumber \\
I_{10} & = & \sum_{i<j,k<l} (m_i^2+m_j^2+m_4^2)\;(n_k^4+n_l^4+n_4^4)
\;G(i,j;k,l), \nonumber \\
I_{11} & = &\sum_{i<j,k<l}(m_i^4+m_j^4+m_4^4)\;(n_k^4+n_l^4+n_4^4)\;G(i,j;k,l),
\label{eq:invexpr} \end{eqnarray}
with
\begin{eqnarray}
G(i,j;k,l) & = & -(m_j^2-m_i^2)(m_4^2-m_i^2)(m_4^2-m_j^2) 
 (n_l^2-n_k^2)(n_4^2-n_k^2)(n_4^2-n_l^2) \nonumber \\ 
& \times & {\mathrm Im\;} (V_{ik} V_{jk}^* V_{jl} V_{il}^*). 
\end{eqnarray}
$I_{4,8}$ involve in principle the imaginary parts of products of six
CKM matrix elements. However, all of these (96) can be written in terms
of only one of them plus products of  ${\mathrm Im\;}(V_{ik} V_{jk}^* 
V_{jl} V_{il}^*)$ times the modulus of a CKM matrix element squared.
 Although the term with the imaginary part of six
$V_{CKM}$ matrix elements cancels in $I_{4,8}$, their expressions are
still too long to be written here. We could also try to write 
${\mathrm Im\;} (V_{ik} V_{jk}^* V_{jl} V_{il}^*)$
and then $I_\alpha$ as a function of the 3 CP violating
phases parametrizing the CKM matrix in the four family case. With
this purpose we could use for instance the parametrization of $V_{CKM}$
in Ref. \cite{papiro11}. However, due to the relatively complicated
dependence of $V_{ij}$ on these 3 phases the expressions for 
$I_\alpha$ are too long to write them here. 
${\mathrm Im\;}(V_{ik} V_{jk}^* V_{jl} V_{il}^*)=0$; $i<j$, $k<l$;
$i,j,k,l=1,2,3$
form also a set of necessary and sufficient constraints for CP
conservation for four quark generations \cite{papiro12}. These
conditions are invariant only under quark eigenstate phase
redefinitions. The cancellation of these 9 quantities
guarantees that $V_{CKM}$ can be made real and viceversa. Unitarity
allows for other choices of 9 conditions, for instance involving
also the fourth family. We conventionally choose not to do so at
present. Thus Eqs. (\ref{eq:invexpr}) prove that $I_{1-3,5-7,9-11}$
are necessary and sufficient constraints for CP conservation in the
case of nondegenerate fermion masses, because they are linear in 
${\mathrm Im\;}(V_{ik} V_{jk}^* V_{jl} V_{il}^*)$ and independent.
Now points ({\em i\/}), ({\em ii\/}) can be answered: CP
is conserved if there are three up (down) degenerate masses,
three pairs of degenerate masses or the 9
quantities  ${\mathrm Im\;}(V_{ik} V_{jk}^* V_{jl} V_{il}^*)$ are
zero. (In the degenerate case one also requires
$I_{4,8}=0$. )

\section{CP violating effects of a fourth family.} 
\begin{figure}[htb]
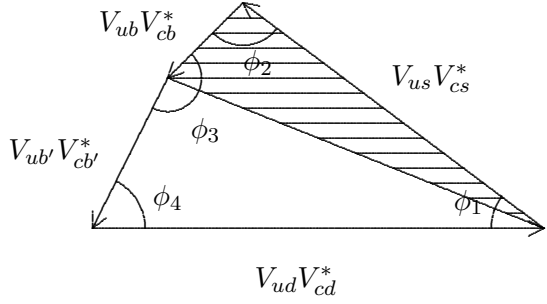

\begin{center}
\mbox{\beginpicture
\setcoordinatesystem units <1cm,1cm>
\unitlength=1cm
\linethickness=1pt
\setplotsymbol ({\thinlinefont .})
\plot 0 0 6 0 /
\plot 6 0 2 3 /
\plot 2 3 1 2 /
\plot 1 2 0 0 /
\plot 1 2 6 0 /
\plot 5.8 0.1 6 0 5.8 -0.1 /
\plot 2.22 2.96 2 3 2.1 2.8 /
\plot 1.0707 2.212 1 2 1.212 2.0707 /
\plot 0 0.2236 0 0 0.179 0.134 /
\circulararc 63.435 degrees from 0.7 0 center at 0 0
\circulararc -36.87 degrees from 5.3 0 center at 6 0
\circulararc 161.56 degrees from 0.8 1.6 center at 1 2
\circulararc 98.13 degrees from 1.6 2.6 center at 2 3
\put{$V_{ud} V_{cd}^*$} at 2.7 -0.7
\put{$V_{ub'} V_{cb'}^*$} at -0.5 1
\put{$V_{ub} V_{cb}^*$} at 0.6 2.7
\put{$V_{us} V_{cs}^*$} at 4.5 2
\put{$\phi_4$} at 1.0 0.4
\put{$\phi_1$} [lB] at 4.8 0.25
\put{$\phi_3$} at 1.5 1.3
\put{$\phi_2$} at 2.2 2.2
\linethickness=0.25pt
\plot 1.8 2.8 2.266 2.8 / 
\plot 1.6 2.6 2.533 2.6 / 
\plot 1.4 2.4 2.8 2.4 / 
\plot 1.2 2.2 3.066 2.2 / 
\plot 1.0 2.0 3.333 2.0 / 
\plot 1.5 1.8 3.6 1.8 / 
\plot 2.0 1.6 3.866 1.6 /
\plot 2.5 1.4 4.133 1.4 / 
\plot 3.0 1.2 4.4 1.2 / 
\plot 3.5 1.0 4.666 1.0 / 
\plot 4.0 0.8 4.933 0.8 / 
\plot 4.5 0.6 5.2 0.6 / 
\plot 5.0 0.4 5.466 0.4 / 
\plot 5.5 0.2 5.733 0.2 / 
\linethickness=0pt
\putrectangle corners at -1 -1 and 7 4
\endpicture}
\end{center}
\caption{Unitarity quadrangle \label{figura2} }
\end{figure}

${\mathrm Im\;} V_{ik} V_{jk}^* V_{jl} V_{il}^*$ can be measured in the
mass eigenstate basis and one of them at least must be non-zero if CP 
is violated. This can be summarized with three
quadrangles in the complex plane. The sides of the quadrangles are for
example the products of the elements of the first line of the CKM
matrix times the elements of the second line complex conjugate, whose
sum is zero by unitarity (see Fig. \ref{figura2})
\begin{equation}
V_{ud} V_{cd}^* + V_{us} V_{cs}^* +V_{ub} V_{cb}^* +V_{ub'}
V_{cb'}^*=0.
\end{equation}
The area of the convex quadrangle drawn with these products is
\begin{eqnarray}
A_{uc} & = &\frac{1}{4} \{ 
|{\mathrm Im\;}(V_{ud} V_{cd}^* V_{cs} V_{us}^*)| + 
|{\mathrm Im\;}(V_{us} V_{cs}^* V_{cb} V_{ub}^*)| + 
|{\mathrm Im\;}(V_{ub} V_{cb}^* V_{cb'} V_{ub'}^*)| \nonumber \\
& + & |{\mathrm Im\;}(V_{ub'} V_{cb'}^* V_{cd} V_{ud}^*)| \} , \label{eq:cuad}
\end{eqnarray}
whereas the angles are the arguments of the four-products,
\begin{eqnarray*}
\sin \phi_1 & = & |\sin \arg\,(V_{ud} V_{cd}^* V_{cs} V_{us}^*)|, \\
\sin \phi_2 & = & |\sin \arg\,(V_{us} V_{cs}^* V_{cb} V_{ub}^*)|, \\
\sin \phi_3 & = & |\sin \arg\,(V_{ub} V_{cb}^* V_{cb'}V_{ub'}^*)|, \\
\sin \phi_4 & = & |\sin \arg\, (V_{ub'} V_{cb'}^* V_{cd} V_{ud}^*)|.
\end{eqnarray*}
Hence the vanishing of the
area of the quadrangle means that the four terms in Eq. (\ref{eq:cuad})
vanish and viceversa. One can consider also the quadrangles and areas
resulting from multiplying the first and third rows or the second and
third ones. Thus the vanishing of $A_{uc,ut,ct}$ means that the
corresponding 12 terms on the right-hand side vanish, and it is easy
to convince oneself using the unitarity of the CKM matrix that this is
equivalent to the cancellation of the 9 independent quantities 
${\mathrm Im\;}(V_{ik} V_{jk}^* V_{jl} V_{il}^*)$ involving only the
first three families and therefore to CP invariance \cite{papiro12}.
A numerical update
of the present limits for a fourth family will be presented elsewhere.
For illustration purposes, however, let us assume that the thesis in
Ref \cite{papiro10} holds and
\begin{equation}
|V_{CKM}| \sim \left( \begin{array}{cccc}
1 & \lambda & \lambda^3 & \lambda^2 \\
\lambda & 1 & \lambda^2 & \lambda \\
\lambda^3 & \lambda^2 & 1 & \lambda \\
\lambda^2 & \lambda & \lambda & 1
\end{array} \right)
\end{equation}
with $\lambda = |V_{us}|=0.22$ \cite{papiro13}. In this case the quadrangles have
areas $A_{uc} \sim \lambda^4$, $A_{ut} \sim \lambda^6$,
$A_{ct} \sim \lambda^4$, respectively. If the fourth family does not
mix, $V_{ib'}=V_{t'j}=0$, $i=u,c,t$; $j=d,s,b$, the three quadrangles
collapse to three triangles (shaded region of the quadrangle in 
Fig. \ref{figura2}) with the same area $A \sim \lambda^6$ in Eq.
(\ref{eq:1:4}).
\appendix
\section{Appendix.}
$H_u$ is diagonal, $(H_u)_{ij}=m_i^2 \delta_{ij}$, with $m_i$ the mass
of the up quark $i$, and $H_d$ hermitian with $(H_d)_{ij}=n_{ij}$,
$n_{ij}=n_{ji}^*$. Let us define $d_{ij}\equiv m_j^2-m_i^2$, $C(i,j,k)
\equiv d_{ij} d_{jk} d_{ki}\;{\mathrm Im\;} n_{ij} n_{jk} n_{ki}$,
then
\begin{eqnarray}
I_1 & = & \sum_{i<j<k} C(i,j,k),  \\
I_2 & = & \sum_{i<j<k} (m_i^2+m_j^2+m_k^2)\; C(i,j,k), \nonumber \\
I_3 & = & \sum_{i<j<k} (m_i^4+m_j^4+m_k^4)\; C(i,j,k), \nonumber \\
I_4 & = & \sum_{i<j<k} (m_i^6+m_j^6+m_k^6)\; C(i,j,k). \nonumber
\end{eqnarray}
The determinant of the coefficients of the C's is just $\prod_{i<j} 
(m_j^2-m_i^2)$, so in the case of nondegenerate quark masses $I_{1-4}=0$ imply
$C(i,j,k)=0$ and all the three-cycles $n_{ij} n_{jk} n_{ki}$
are real. However, there are still
CP violating solutions with $n_{ij}=n_{kl}=0$, where $i,j,k,l$ are four
distinct indices. In Ref. \cite{papiro6}, this is identified as the only case in
which the reality of the three-cycles does not imply the reality of the
four-cycles $n_{ij} n_{jk} n_{kl} n_{li}$. Due to the 
symmetry of the problem, we can assume $n_{12}=n_{34}=0$. In this case
all the
four-cycles but $n_{13} n_{32} n_{24} n_{41}$ are real. If 
$n_{12}=n_{34}=0$,
\begin{eqnarray*}
{I_5} & = &d_{12}\, d_{34}\, (m_1+m_2-m_3-m_4)\; {\mathrm Im\;} 
n_{13} n_{32} n_{24} n_{41},  \\
{I_6} & = & d_{12}\, d_{34}\, (m_1^2+m_1 m_2+m_2^2-m_3^2-m_3 m_4-m_4^2)\; 
{\mathrm Im\;} n_{13} n_{32} n_{24} n_{41},
\end{eqnarray*}
then $I_{5,6}=0$ imply ${\mathrm Im\;} n_{13} n_{32} n_{24} n_{41}=0$ and 
CP conservation.

In the case of degenerate masses we can assume without loss of
generality $m_1^2=m_2^2$, $n_{12}=0$. In this case $I_{2-4}$ are
proportional to $I_1$,
\begin{eqnarray*}
I_5 & = & (n_{11}+n_{33}+n_{44})\; C(1,3,4) + (n_{22}+n_{33}+n_{44})\; C(2,3,4)
\end{eqnarray*}
and $I_6$ is a linear combination of
$I_{1,5}$. We can distinguish two subcases. If $n_{11}=n_{22}$, we can transform
$H_d$ and assume $n_{23}=0$. Then $I_1=0$ guarantees CP conservation. If 
$n_{11}\neq n_{22}$, $I_1$ and $I_5$ still have two
CP violating solutions corresponding to $n_{34}=0$ and $m_3^2=m_4^2$. In the
latter case we can transform $H_d$ and also assume $n_{34}=0$. If $m_1=m_2$,
$n_{12}=n_{34}=0$,
\begin{eqnarray*}
I_7 & = & d_{13}\, d_{34}\, d_{41}\, (n_{11}-n_{22}) \; {\mathrm Im\;} 
n_{13} n_{32} n_{24} n_{41}, \\
I_8 & = & d_{13}\, d_{41}\, (n_{11}-n_{22}) (d_{14} n_{44}-d_{13} 
n_{33})\;{\mathrm Im\;} n_{13} n_{32} n_{24} n_{41},
\end{eqnarray*}
and $I_{7,8}=0$ ensure CP conservation.

\vspace{1cm}
\noindent
{\Large \bf Acknowledgements}

\vspace{0.4cm}
\noindent
We thank G. Branco and M. Zra{\l}ek for discussions. This work was
partially supported by CICYT under contract AEN94-0936, by the Junta
de Andaluc\'{\i}a and by the European Union under contract
CHRX-CT92-0004.


\begin{thebibliography}{99}
\bibitem{papiro1}
N. Cabibbo, Phys. Rev. Lett. {\bf 10}, 531 (1963);
M. Kobayashi and T. Maskawa, Prog. Theor. Phys. {\bf 49}, 652 (1973)
\bibitem{nose1}
F. D. Murnaghan, {\em The Unitary and Rotation Groups}, Spartan, Washington D.
C., 1962
\bibitem{papiro1a}
The LEP Collaborations ALEPH, DELPHI, L3, OPAL and the LEP Electroweak
Working Group, CERN preprint, CERN-PPE/95-172
\bibitem{papiro1b}
Review of Particle Properties, Phys. Rev. {\bf D50} (1994)
\bibitem{papiro1c}
A. Datta and E. A. Paschos, in {\em CP violation}, edited by C. Jarlskog (World
Scientific, Singapore, 1989) 292, and references there in; Proceedings of the
2nd International Symposium on The Fourth Family of Quarks
and Leptons, edited by D. B. Cline and A. Soni, Santa Monica (1989)
\bibitem{papiro2}
C. Jarlskog, Phys. Rev. Lett. {\bf 55}, 1039 (1985); Z. Phys. {\bf C29}, 491
(1985)
\bibitem{papiro3}
J. Bernab\'{e}u, G. C. Branco and M. Gronau, Phys. Lett. {\bf 169B},
243 (1986); for an invariant formulation of Higgs couplings see L. Lavoura and
J. P. Silva, Phys. Rev. {\bf D50}, 4619 (1994); F. J. Botella and J. P. Silva,
Phys. Rev. {\bf D51}, 3870 (1995)
\bibitem{papiro4}
L.-L. Chau and W.-Y. Keung, Phys. Rev. Lett. {\bf 53}, 1802 (1984)
\bibitem{papiro8}
C. Jarlskog and R. Stora, Phys. Lett. {\bf 208B}, 268 (1988)
\bibitem{papiro9}
See for instance A. Ali and D. London, DESY-95-148, hep-ph/9508272; 
A. J. Buras, MPI-PHT-94-30, hep-ph/9406272
\bibitem{papiro6}
M. Gronau, A. Kfir and R. Loewy, Phys. Rev. Lett. {\bf 56}, 1538 (1986);
see also O. W. Greenberg, Phys. Rev. {\bf D32}, 1841 (1985); D. D. Wu, Phys.
Rev. {\bf D33}, 860 (1986)
\bibitem{papiro7}
S. Wolfram, Mathematica, a System for Doing Mathematics by Computer
(Addison-Wesley Publishing Company, Redwood City, California, 1988)
\bibitem{papiro7b}
F. del Aguila, J. A. Aguilar-Saavedra, M. Zra{\l}ek, UG-FT-63/96,
hep-ph/9607311
\bibitem{papiro11}
V. Barger, K. Whisnant and R. J. N. Phillips, Phys. Rev. {\bf D23}, 2773 (1981)
\bibitem{papiro12}
F. J. Botella and L.-L. Chau, Phys. Lett. {\bf 168B}, 97 (1986); J. D. Bjorken
and I. Dunietz, Phys. Rev. {\bf D36}, 2109 (1987) 
\bibitem{papiro10}
C. Hamzaoui, A. I. Sanda and A. Soni, Phys. Rev. Lett. {\bf 63}, 128
(1989); see also M. Gronau and J. Schechter, Phys. Rev. {\bf D31}, 1668 (1985);
X.-G. He and S. Pakvasa, Nucl. Phys. {\bf B278}, 905 (1986)
\bibitem{papiro13}
L. Wolfenstein, Phys. Rev. Lett. {\bf 51}, 1945 (1983)
\end{thebibliography}
\end{document}